\documentclass[journal]{IEEEtran}
\IEEEoverridecommandlockouts
\usepackage{cite}
\usepackage{amsmath,amssymb,amsfonts}
\usepackage{algorithmic}
\usepackage{graphicx}
\usepackage{textcomp}
\usepackage{xcolor}
\usepackage{amsmath}
\usepackage{physics}
\usepackage{amssymb}
\usepackage{verbatim}
\usepackage{mathtools}
\usepackage{hyperref}
\usepackage{subcaption}
\usepackage{bbm}

\def\BibTeX{{\rm B\kern-.05em{\sc i\kern-.025em b}\kern-.08em
    T\kern-.1667em\lower.7ex\hbox{E}\kern-.125emX}}
    
\newtheorem{theorem}{Theorem}[section]
\newtheorem{corollary}[theorem]{Corollary}
\newtheorem{lemma}[theorem]{Lemma}

\newtheorem{remark}{Remark}

\begin{document}

\title{Secrecy Capacity of a Gaussian Wiretap Channel With ADCs is Always Positive}

\author{Seung-Hyun Nam,~\IEEEmembership{Student Member,~IEEE,}
        and Si-Hyeon Lee,~\IEEEmembership{Member,~IEEE}
\thanks{S.-H. Nam and S.-H. Lee are with the School of Electrical Engineering, Korea Advanced Institute of Science and Technology (KAIST), Daejeon, South Korea (e-mail: shnam@kaist.ac.kr, sihyeon@kaist.ac.kr). This paper was presented in part at IEEE ITW 2019, Visby, Sweden \cite{Cs_real_onebit}. A partial error in \cite{Cs_real_onebit}, i.e., Lemma 4.1 of \cite{Cs_real_onebit} does not hold in general,  has been corrected in this paper. }
}

\maketitle

    \begin{abstract}
        We consider a complex Gaussian wiretap channel with finite-resolution analog-to-digital converters (ADCs) at both the legitimate receiver and the eavesdropper.
        For this channel, we show that a positive secrecy rate is always achievable as long as the channel gains at the legitimate receiver and at the eavesdropper are different, regardless of the quantization levels of the ADCs.
        For the achievability, we first consider the case of one-bit ADCs at the legitimate receiver and apply a binary input distribution where the two input points have the same phase when the channel gain at the legitimate receiver is less than that at the eavesdropper, and otherwise the opposite phase. Then the result is generalized for the case of arbitrary finite-resolution ADCs at the legitimate receiver by translating the input distribution appropriately. For the special case of the real Gaussian wiretap channel with one-bit ADCs at both the legitimate receiver and the eavesdropper, we show that our choice of input distribution satisfies a necessary condition of optimal distributions for Wyner codes.
        \end{abstract}


\section{Introduction}

The wiretap channel first studied by Wyner \cite{wyner} is a canonical model for physical layer security.
In a wiretap channel, a transmitter wants to send its message reliably to a legitimate receiver while keeping it secret from an eavesdropper.
In this situation, the fundamental limit of the communication rate, the secrecy capacity, has been characterized first for a memoryless degraded wiretap channel \cite{wyner}, and for a general memoryless wiretap channel \cite{secrecycapacity}, in the form of optimization over probability distributions.
If the channel is given precisely, such optimization problem can be solved analytically for some cases.
For the standard Gaussian wiretap channel, the optimization problem for the secrecy capacity was solved exactly, and it was shown that the secrecy capacity is zero when the signal-to-noise ratio (SNR) at the legitimate receiver is less than the SNR at the eavesdropper \cite{gaussianwiretap,Mimowiretap}. 

In practice, the digital wireless communication systems employ analog-to-digital converters (ADCs) at the receivers. If the resolutions of the ADCs are high enough and the wireless channel is modeled as the Gaussian channel, then the digital communication channel can be treated as the ideal Gaussian channel.
But, high resolution ADCs are power-expensive because the power consumption of an ADC increases exponentially in the number of its quantization levels \cite{ADCpower}. Recently, various communication strategies when low-resolution ADCs are employed at the receivers have been studied to enable low-power communications.
For real and complex point-to-point Gaussian channels with one-bit ADCs, the binary phase-shift keying (BPSK) and the rotated quadratic phase-shift keying (QPSK) were shown to achieve the channel capacity, respectively \cite{p2pADC,MIMO_onebit_capacity}. However, the optimal input distribution is not known for the channel with arbitrary ADCs because it is difficult to solve the optimization problem to characterize the capacity analytically.  For wiretap channels with quantizers, \cite{MIMO_DAC} analyzed an asymptotic achievable downlink secrecy rate for a MIMO wiretap channel with digital-to-analog converters (DACs) at the base station. Also, for a MIMO wiretap channel where an active eavesdropper tries to spoil the channel estimation at the base station with one-bit ADCs, the downlink secrecy rate was studied in \cite{Downlink_One_bit_MIMO}. The previous works are reviewed in a greater detail in Section~\ref{subsec:prev}. To the best of our knowledge, there have been no  information-theoretic studies on the classical Gaussian wiretap channel with finite-resolution ADCs at the legitimate receiver and the eavesdropper.

In this paper, we consider a complex Gaussian wiretap channel with finite-resolution ADCs at both the legitimate receiver and the eavesdropper, as a model for the low-power physical layer secure communication.
Intuitively, if the channel gain at the eavesdropper is higher than that at the legitimate receiver and the quantization at the eavesdropper is finer than that at the legitimate receiver, one may think that a positive secrecy rate would not be achievable because the eavesdropper observes less distorted signals than the legitimate receiver. Somewhat surprisingly, by exploiting the quantization effect due to the ADCs, we show that a positive secrecy rate is always achievable whenever the channel gains of the legitimate channel and the eavesdropper channel are not equal. This result holds regardless of the resolutions and the thresholds of the ADCs.
To show the achievability of a positive secrecy rate, we first focus on the case of symmetric one-bit ADCs at the legitimate receiver. For such a case,  a binary input distribution is considered where the two input points have the same phase when the channel gain at the legitimate receiver is less than that at the eavesdropper, and otherwise the opposite phase. The resultant achievable secrecy rate is analyzed by approximating to  Z-channels. Then the result is generalized for the case of arbitrary finite-resolution ADCs at the legitimate receiver by translating the input distribution appropriately. 

Furthermore, we partially justify our choice of the input distributions. Our achievability result implies that the channel is not more capable and hence it is not clear whether we can set the auxiliary random variable as the channel input variable in the secrecy capacity expression without loss of optimality. Because it is tricky to handle the auxiliary random variable in general, we consider the maximally achievable rate by the Wyner code in \cite{wyner}. For a real Gaussian wiretap channel with one-bit ADCs, we show that the optimal input distribution for the Wyner code should follow the property of our choice of input distribution, i.e., if the channel gain at the legitimate channel is less than that at the eavesdropper channel, the support of the optimal input distribution should be included in one of the positive or negative regions, and otherwise it should not.

The paper is organized as follows. In Section \ref{sec:problem_formulation}, we formulate our problem, and then review the related works and summarize our contribution. The achievability of a positive secrecy rate is proved in Section \ref{sec:achievability} and a necessary condition for the optimal input distributions for the Wyner code is presented in Section \ref{sec:optimal_support}. We conclude this paper with some discussions in \mbox{Section \ref{sec:discussion}}.

\subsection{Notations}
If the probability mass function (PMF) or the probability density function (PDF) is well defined for a probability distribution $P_X$, we use the notation $P_X$ to denote the corresponding PMF or PDF, and similarly for the conditional distribution $P_{Y|X=x}$. The support of a distribution $P_X$ is denoted as  $\mathcal{S}(P_X)$.
If $X$ follows a distribution $P$, $\mathbb{E}_{X\sim P}$ denotes the expectation with respect to $X$, and the subscript is omitted if it is obvious from the context. 
We denote $\mathbb{E}[X \cdot \mathbbm{1}_{\{X\in A\}}]$ by $\mathbb{E}[X;X\in A]$, where $\mathbbm{1}$ denotes the indicator function.
For given probability distribution $P_X$ of a real random variable $X$, $P_{|X|}$ denotes the probability distribution of $|X|$. 
For random variables $X,Y,$ and $Z$, ${X-Y-Z}$ denotes a Markov chain, i.e., $X$ and $Z$ are conditionally independent given $Y$.
The binary entropy function is denoted as $h(\cdot)$, and $f_{\mathcal{N}}(x;\mu,\sigma^2)$ is the PDF of the Gaussian distribution with mean $\mu$ and variance $\sigma^2$.
The function $Q(\cdot)$ is the tail distribution function of the standard normal distribution, 
\begin{equation}
    Q(x) = \int_x^\infty f_{\mathcal{N}}(u;0,1) du. \nonumber
\end{equation}
The sign function, $\mathrm{sgn}(\cdot)$, refers
\begin{equation}
    \mathrm{sgn}(x) = \begin{cases} \;\:\: 1, & x \geq 0 \\ -1, & x < 0 \end{cases}, \nonumber
\end{equation}
and $[x]^+=\max\{x,0\}$. For $x \in \mathbb{C}$, $\mathcal{R}(x)$ (resp. $\mathcal{I}(x)$) denotes the real (resp. imaginary) part of $x$ and $j$ denotes $\sqrt{-1}$.
The sets $\{x\in \mathbb{R}:x\geq 0\}$ and $\{x\in \mathbb{R}:x \leq 0\}$ are denoted as $\mathbb{R}_+$ and $\mathbb{R}_-$, respectively. For integers $a$ and $b$, $[a:b]$ denotes the set $\{a,a+1,\cdots,b-1,b\}$.

\section{Problem Formulation}\label{sec:problem_formulation}
In Section \ref{sec:achievability}, the positivity of secrecy capacity is shown first for the case with one-bit ADCs at the legitimate receiver. Then, the results are  generalized to the case with arbitrary finite-resolution ADCs at the legitimate receiver through  some simple manipulation. Hence, for simplicity, we formulate the problem for the channel with one-bit ADCs at the legitimate receiver. 
 
\subsection{Model}

\begin{figure}[t]
    \centering
    \includegraphics[width=0.8\columnwidth]{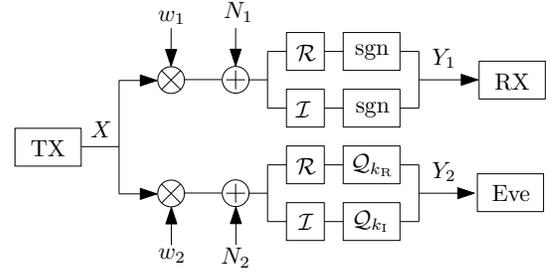}
    \caption{A Gaussian wiretap channel with one-bit ADCs at the legitimate receiver and finite-resolution ADCs at the eavesdropper.}
    \label{fig:Model}
\end{figure}

Consider a memoryless complex Gaussian wiretap channel with one-bit ADCs at the legitimate receiver and finite-resolution ADCs at the eavesdropper. The quantization is assumed to be applied separately for real and imaginary parts. 
For the eavesdropper, we assume $k_{\mathrm{R}}\geq 2$ and $k_{\mathrm{I}}\geq 2$ quantization points for the real and imaginary parts, respectively. The quantization function of the $k_i$-point ADC $\mathcal{Q}_{k_i}(\cdot)$ for $i \in \{\mathrm{R},\mathrm{I}\}$ is given as
\begin{equation}
    \mathcal{Q}_{k_i}(x) = y_{i,l} \text{ if }x \in [q_{i,l-1},q_{i,l}), 
\end{equation}
for all $l \in [1:k_i]$, where  ${(q_{i,1},\cdots,q_{i,k_i -1}) \in \mathbb{R}^{k_i-1}}$ are the threshold points, ${(y_{i,1},\cdots,y_{i,k_i}) \in \mathbb{R}^{k_i}}$ are the output points, and ${q_{i,0}=-\infty}$, ${q_{i,{k_i}}=\infty}$. The threshold points are assumed to be ${q_{i,l-1}<q_{i,l}}$ for all $l \in [1:k_i]$, and output points are distinct. For the ADCs at the legitimate receiver, we consider one-bit ADCs, and an one-bit ADC  corresponds to the quantization function with only one threshold point $0$, and the output points $(-1,1)$. 

The transmitter sends a channel input $X$, then the legitimate receiver and the eavesdropper observe $Y_1$ and $Y_2$, respectively, which follows the input-output relationship given as 
\begin{multline}
    Y_1 = \textrm{sgn}(\mathcal{R}(w_1 X+N_1)) + j \cdot \textrm{sgn}(\mathcal{I}(w_1 X+N_1)),
\end{multline}
\begin{multline}
    Y_2 = \mathcal{Q}_{k_{\mathrm{R}}} (\mathcal{R}(w_2 X+N_2)) + j \cdot \mathcal{Q}_{k_{\mathrm{I}}} (\mathcal{I}(w_2 X+N_2)),
\end{multline}
as depicted in Fig.~\ref{fig:Model}.
The complex channel gains $w_1$, $w_2$ are assumed to be non-zero constants, and known at both the transmitter and the legitimate receiver. The Gaussian noises $N_i \sim \mathcal{CN}(0,2)$ are independent with each other and with $X$.
Thus, the channel transition probabilities are given as
\begin{align}
    P_{Y_1|X}( 1 \pm j|x) &= Q\left( - \mathcal{R}(w_1 x) \right) \cdot Q\left( \mp \mathcal{I}(w_1 x) \right), \label{eq:tran_p_11}
    \\ P_{Y_1|X}(- 1 \mp j|x) &= Q\left(  \mathcal{R}(w_1 x) \right) \cdot Q\left( \pm \mathcal{I}(w_1 x) \right),\label{eq:tran_p_12}
\end{align}
\begin{align}
    &P_{Y_2|X}(y_{\mathrm{R},i}+j\cdot y_{\mathrm{I},l}|x) \nonumber
    \\& =\left(Q\left( q_{\mathrm{R},i-1}-\mathcal{R}(w_2 x) \right) - Q\left( q_{\mathrm{R},i}-\mathcal{R}(w_2 x) \right) \right) \label{eq:tran_p_2}
    \\& \quad\quad \cdot \left(Q\left( q_{\mathrm{I},l-1}-\mathcal{I}(w_2 x) \right) - Q\left( q_{\mathrm{I},l}-\mathcal{I}(w_2 x) \right) \right), \nonumber
\end{align}
for $i \in [1:k_{\mathrm{R}}]$, and $l \in [1:k_{\mathrm{I}}]$.

Through $n \in \mathbb{N}$ channel uses, the transmitter encodes a uniformly distributed message $M \in \mathcal{M} = \left[ 1:2^{\lceil nR \rceil} \right]$ into the channel input $X^n=f_n (M)$ with some encoding function $f_n:\mathcal{M} \rightarrow \mathbb{C}^n$ satisfying the average power constraint $J>0$, i.e., 
\begin{equation}
    \mathbb{E}\left[\frac{1}{n}\sum_{i=1}^n |X_i|^2 \right] \leq J. \label{eq:powerconstraint}
\end{equation}
The legitimate receiver decodes  $\hat{M}=g_n(Y_1^n)$ based on the observation $Y_1^n$ using a decoding function $g_n:\mathbb{C}^n \rightarrow \mathcal{M}$.
As in \cite{wyner}, a secrecy rate $R$ is said to be achievable if there exists a sequence of $\{(f_n,g_n)\}_{n \in \mathbb{N}}$ satisfying
\begin{equation}
    \lim\limits_{n \rightarrow \infty} \mathbb{P} \{\hat{M} \neq M\}=0,\; \lim\limits_{n \rightarrow \infty}\frac{1}{n} I(M;Y_2^n) =0.
\end{equation}
The secrecy capacity $C_s$ is defined as the supremum of achievable secrecy rates.

\subsection{Previous Work and Our Contribution} \label{subsec:prev}
In the following, we review some of previous works on 1) wiretap channels (without ADCs) and 2) (non-wiretap) channels with ADCs. Then the main contribution of this paper is summarized. 
\subsubsection{Wiretap Channel}
The wiretap channel was first studied by Wyner in \cite{wyner}, where the secrecy capacity of a degraded wiretap channel was characterized as
\begin{equation}\label{eq:Degraded_Cs}
    C^{\text{D}}_s = \sup\limits_{P_X} I(X;Y_1)-I(X;Y_2).
\end{equation}
For a general wiretap channel with the average power constraint $J$, \cite{secrecycapacity} showed that the secrecy capacity is given as
\begin{equation} \label{eq:Cs}
    C_s = \sup\limits_{P_{U,X}:\substack{U-X-(Y_1,Y_2) \\ \mathbb{E}[X^2] \leq J}} I(U;Y_1) - I(U;Y_2).
\end{equation}
We note that it is sufficient to set $U=X$ in \eqref{eq:Cs} for a class of more capable channels where $I(X;Y_1) \geq I(X;Y_2)$ for all $P_X$, which includes the degraded channels \cite{secrecycapacity}. 

If the legitimate channel and the eavesdropper channel are Gaussian \cite{gaussianwiretap,Mimowiretap} with average power constraint $J$, the secrecy capacity is simplified to
\begin{equation}
    C_s^{\text{G}}
    = \left[ \log\left(1+ \frac{|w_1|^2 J}{2} \right)
     - \log\left( 1+ \frac{|w_2|^2 J}{2} \right) \right]^+,
\end{equation}
which is achieved by letting $U=X$ and $X \sim \mathcal{CN}(0,J)$ in $\eqref{eq:Cs}$. This follows from the fact that the Gaussian wiretap channel is degraded. Hence, a positive secrecy rate of the Gaussian wiretap channel without ADCs is not achievable when $|w_1|\leq |w_2|$, which makes sense because the eavesdropper observes a signal with a better quality.

\subsubsection{Channels with ADCs}
For a real Gaussian channel with average power constraint $J$ where the receiver employs a one-bit ADC, \cite{p2pADC} showed that the BPSK with power $J$ achieves the capacity of 
\begin{equation}
    C_{\mathbb{R},1\text{-bit}} = 1-h\left(Q\left(|w|\sqrt{J}\right)\right).
\end{equation}
Moreover, \cite{p2pADC} showed that the capacity for the channel with a $k$-point ADC at the receiver can be achieved by a discrete distribution with at most $k+1$ points of support, by applying Karush-Kuhn-Tucker (KKT) conditions.
For a complex Gaussian channel with component-wise one-bit ADCs at the receiver, \cite{MIMO_onebit_capacity} showed that the  capacity is given as 
\begin{equation} 
    C_{\mathbb{C},\text{1-bit}} = 2\left(1-h\left(Q\left(|w|\sqrt{\frac{J}{2}}\right)\right)\right), \label{eq:p2p_complex_capacity}
\end{equation}
and the QPSK with the phase rotation $-\angle w$ and power $J$ achieves the capacity.\footnote{In this paper, $\sqrt{J/2}$ is in \eqref{eq:p2p_complex_capacity} instead of $\sqrt{J}$, because we set the variance of complex Gaussian noise to 2.}
For the aforementioned real and complex channels with one-bit ADCs, the optimal input distribution was analyzed by exploiting the concavity of the mutual information $I(X;Y)$ in $P_X$, the symmetry of the channel, and the convexity of $h(Q(\sqrt{\cdot}))$ \cite{hqsqrt}.

\subsubsection{Our Contribution}

First, we show that a positive secrecy rate is achievable whenever $|w_1| \neq |w_2|$, no matter what the thresholds of ADCs are. To show the achievability, we focus on the Wyner code  \cite{wyner} which achieves $I(X;Y_1)-I(X;Y_2)$ for input distribution $P_X$. 
One might expect that if the one-bit ADCs are at the legitimate receiver, $|w_1|>|w_2|$, and $P_X$ is set to the rotated QPSK, which maximizes $I(X;Y_1)$, then it would be easily shown that a positive secrecy rate is achievable. 
However, it is tricky to handle $I(X;Y_2)$ exactly or find a tight upper bound on it  as the number of possible realizations of $Y_2$ is $k_{\mathrm{R}} \cdot k_{\mathrm{I}}$. Thus, it is not clear whether such QPSK achieves a positive secrecy rate in general (if  one-bit ADCs are also employed at the eavesdropper, it can be proved that a positive secrecy rate is achievable, which is proved in Appendix A).
As a way to avoid this difficulty, we consider a binary input distribution and analyze the resultant rate by approximating each of the legitimate and the eavesdropper channels to a Z-channel.

Second, for a real Gaussian channel with one-bit ADCs, we find a necessary condition for the optimal distributions for Wyner code. In contrast to a Gaussian wiretap channel without ADCs, the sufficiency of $U=X$ in \eqref{eq:Cs} is not straightforward because our channel is shown to be not capable.
Because it  is  tricky  to  handle  the auxiliary  random  variable $U$ in general, we consider the maximally achievable secrecy rate by the Wyner code given in \eqref{eq:Degraded_Cs}, which is also used for showing the positivity of the secrecy capacity. 
The Wyner code $\eqref{eq:Degraded_Cs}$ is of practical interest, because there is a polar code for wiretap channels \cite{MV11} which achieves the secrecy rate of $I(X;Y_1)-I(X;Y_2)$ when $P_X$ is set to a binary uniform  distribution.\footnote{The channel was restricted to a symmetric channel in \cite{MV11}, but it can be checked that the secrecy rate of $I(X;Y_1)-I(X;Y_2)$ is also achievable for any binary-input memoryless discrete wiretap channel.}
Also, a study in \cite{cyclic_symmetry} shows that the characterization of  \eqref{eq:Degraded_Cs} can be used to find the secrecy capacity \eqref{eq:Cs} (a detailed discussion is in Section~\ref{sec:discussion}).
    
Even if $U=X$, the optimal distributions cannot be found directly by the previous techniques. For our Gaussian wiretap channel with ADCs, $I(X;Y_1)-I(X;Y_2)$ is not concave in $P_X$. Therefore, we cannot use the previous techniques \cite{p2pADC} directly for finding the optimal distributions for $\eqref{eq:Degraded_Cs}$. Moreover, because the difference between two mutual information terms is optimized in \eqref{eq:Degraded_Cs}, the technique used for proving the sufficiency of finite support in \cite{p2pADC}, which relies on some monotonic property related to  a single mutual information term, cannot be applied directly.

\section{Achievability of a Positive Secrecy Rate} \label{sec:achievability}
In this section, we show that a positive secrecy rate is achievable regardless of the quantization levels of the ADCs as long as $|w_1| \neq |w_2|$. 

In the following, we focus on the achievability of a positive secrecy rate without a power constraint. If a positive secrecy rate is achievable by using possibly very large (but finite) power, this implies that it is also achievable in the presence of the power constraint since the transmit power can be adjusted to satisfy the power constraint by time-burst transmission, i.e., use the scheme for ${0< \alpha \leq J/\mathbb{E}[|X|^2]}$ fraction of time and stay idle for the remaining time. 

For a channel without power constraint, the Wyner code \cite{wyner} with input distribution $P_X$ achieves the secrecy rate $R_s(P_X)$ given as 
\begin{multline}
    R_s(P_X) := I(P_X,P_{Y_1|X})-I(P_X,P_{Y_2|X})
    \\= I(X;Y_1)-I(X;Y_2).
\end{multline}

The following theorem states that $R_s(P_X)>0$ for some $P_X$ for the case with one-bit ADCs at the legitimate receiver.
\begin{theorem}
For a Gaussian wiretap channel with one-bit ADCs at the legitimate receiver and finite-resolution ADCs at the eavesdropper, there exists $P_X$ such that $R_s(P_X)>0$ and $\mathbb{E}[|X|^2] < \infty$ whenever $|w_1|\neq |w_2|$.  \label{thm_achievability}
\end{theorem}

This theorem is generalized to the channel with arbitrary finite-resolution ADCs at the legitimate receiver in \mbox{Corollary \ref{cor:general}}. 

Before a precise and rigorous proof, let us first present the main ideas and intuitions used in the proof. 
In general, it is tricky to handle the mutual information $I(X;Y_2)$ exactly, because it involves a number of terms as the number of possible realizations of $Y_2$ is $k_{\mathrm{R}} \cdot k_{\mathrm{I}}$. 
To avoid this difficulty, we choose $P_X$ as a binary distribution whose support contains a point with very large absolute value.
For such $P_X$, each of the legitimate channel and the eavesdropper channel can be approximated to Z-channels. Since the mutual information for a Z-channel is decreasing in the crossover probability, it is sufficient to compare only the crossover probabilities of each equivalent Z-channels to show $R_s(P_X)>0$.

\begin{figure}[t]
    \centering
    \includegraphics[width=0.8\columnwidth]{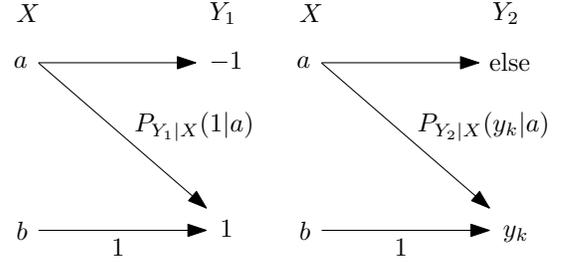}
    \caption{The Z-channels equivalent to the original channels when $b \rightarrow \infty$.}
    \label{fig:zchannel}
\end{figure}

\begin{figure}
    \centering
    
    \begin{subfigure}[b]{\columnwidth}
    \centering
    \includegraphics[width=0.85\columnwidth]{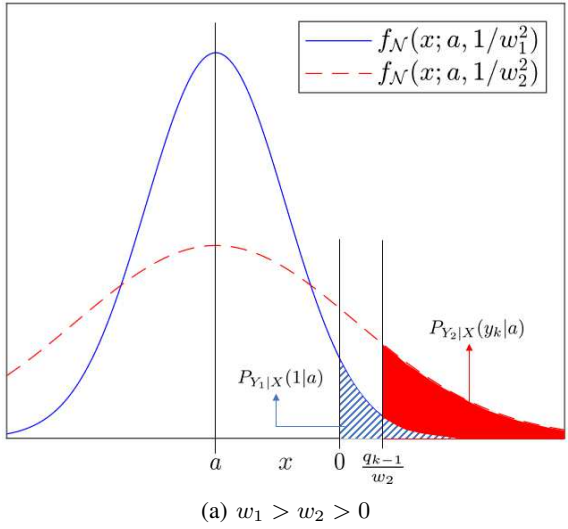}
    \caption{$w_1 > w_2 > 0$  \\ \vspace{10pt}}
    \end{subfigure}
    
    \begin{subfigure}[b]{\columnwidth}
    \centering
    \includegraphics[width=0.85\columnwidth]{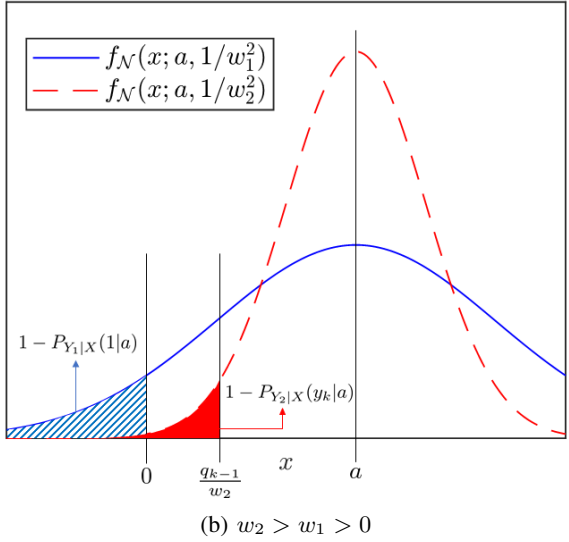}
    \caption{$w_2>w_1>0$}
    \end{subfigure}

    \caption{For given any given $(w_1,w_2,q_{k-1})$, $w_1\neq w_2$, there exists $a$ which satisfies $P_{Y_1|X}(1|a) < P_{Y_2|X}(y_k|a)$ because the tail of the PDF of a Gaussian distribution is steeper as its variance increases.}
    \label{fig:Gaussian_Intuition}
\end{figure}

To illustrate this intuition, assume that the channel is a real (instead of complex) channel where the one-bit ADC is at the legitimate receiver and the $k$-point ADC is at the eavesdropper. Suppose $w_1,w_2>0$, and let $P_X$ be a binary distribution given as 
\begin{equation}
    P_X(x) = \begin{cases} \phi & \text{ if } x=a \\ 1-\phi & \text{ if } x=b \end{cases}, \label{eq:binary_dist_real}
\end{equation}
where $0<\phi<1$, and $b$ is sufficiently large.
Then, both $P_{Y_1|X}(1|b)$ and $P_{Y_2|X}(y_k|b)$ are close to $1$, because the Gaussian noises can be ignored for large $b$.
Equivalently, $P_{Y_1|X}(-1|b)$ and $P_{Y_2|X}(y_l|b)$ are close to $0$ for all $l<k$.
Therefore, the conditional entropies are approximated to 
\begin{align}
    & H(X|Y_1) \approx P_{Y_1}(1)\cdot H(X|Y_1=1), \label{eq:condent1}
    \\& H(X|Y_2) \approx P_{Y_2}(y_k)\cdot H(X|Y_2=y_k). \label{eq:condent2}
\end{align}
The RHS of \eqref{eq:condent1} and \eqref{eq:condent2} are equal to the corresponding conditional entropies of Z-channels depicted in Fig. \ref{fig:zchannel}. Hence, the original channels can be regarded as the Z-channels because the mutual informations are preserved. Therefore, if there exists $a$ which satisfies $P_{Y_1|X}(1|a) < P_{Y_2|X}(y_k|a)$, then $R_s(P_X)>0$ can be achieved by choosing such $a$ in \eqref{eq:binary_dist_real}. 
The existence of such $a$ can be checked graphically in Fig. \ref{fig:Gaussian_Intuition}. 
Consider first the case of $w_1>w_2$. In \mbox{Fig. \ref{fig:Gaussian_Intuition}-(a)}, the blue dashed area and the red solid area correspond to $P_{Y_1|X}(1|a)$ and $P_{Y_2|X}(y_k|a)$, respectively. We can show that there exists $a<0$ such that a larger variance of $1/w_2^2$ overcome the effect of constant gap $q_{k-1}/w_2$ of thresholds so that $P_{Y_2|X}(y_k|a)>P_{Y_1|X}(1|a)$. 
For the case $w_1<w_2$, in Fig. \ref{fig:Gaussian_Intuition}-(b),  the blue dashed area and the red solid area correspond to $1-P_{Y_1|X}(1|a)$ and $1-P_{Y_2|X}(y_k|a)$, respectively. Due to the similar reason, we can show that there exists $a>0$ such that  $1-P_{Y_1|X}(1|a)>1-P_{Y_2|X}(y_k|a)$, i.e., $P_{Y_2|X}(y_k|a)>P_{Y_1|X}(1|a)$.

The intuition from a real channel can be generalized into a complex channel, but the phases of $w_1$ and $w_2$ should be considered when converting to equivalent Z-channels. Let $P_X$ be a binary distribution with phase $\Phi$, i.e., 
\begin{equation}
    P_X(x) = \begin{cases} \phi & \text{if } x=ae^{j\Phi} \\ 1-\phi & \text{if } x=be^{j\Phi}\end{cases},
\end{equation}
where $0<\phi<1$, $a,b \in \mathbb{R}$, and $b$ is sufficiently large.
If $\angle (w_i X) = \Phi+\angle w_i$ is a multiple of $\pi/2$ for some $i=1,2$, then $Y_i$ given $X=be^{j\Phi}$ does not converge to one point as $b$ increases, because $\mathcal{R}(w_ib)$ or $\mathcal{I}(w_ib)$ becomes 0 in \eqref{eq:tran_p_11}-\eqref{eq:tran_p_2}. Therefore, to apply the Z-channel intuition illustrated for the real channel, $\Phi+ \angle w_i$ should not be a multiple of $\pi/2$ for $i=1,2$.

Now, even if $\Phi+ \angle w_i$ is not a multiple of $\pi/2$ so that $Y_i$ converges to one point as $b$ increases for $i=1,2$, comparing the crossover probabilities of equivalent Z-channels is not simple. 
Let $\theta := \angle (w_2 X) = \Phi + \angle w_2$, and ${\Delta := \angle w_1 - \angle w_2}$, so that $\theta + \Delta = \angle (w_1 X)$.
To analyze the equivalent Z-channels, we need to specify the quantized points $\bar{y}_i$ such that $P_{Y_i|X}(\bar{y}_i|be^{j\Phi}) \approx 1$ for sufficiently large $b$.
Because $\bar{y}_1$ and $\bar{y}_2$ depend on $\theta+\Delta$ and $\theta$, respectively, the crossover probabilities $p_i=P_{Y_i|X}(\bar{y}_i|ae^{j\Phi})$ also depend on such phases.
For example, assume $\theta$ (or $\Phi$, equivalently) is set to ${\theta \in (0,\pi/2)}$.
Then, we have ${\bar{y}_2 = y_{\mathrm{R},k_{\mathrm{R}}}+j\cdot y_{\mathrm{I},k_{\mathrm{I}}}}$, and
\begin{multline}
    p_2 = Q\left( q_{\mathrm{R},k_{\mathrm{R}}-1}-|w_2|a\cos\theta \right)
    \\ \cdot Q\left( q_{\mathrm{I},k_{\mathrm{I}}-1}-|w_2|a\sin\theta \right). \label{eqn:p2}
\end{multline}
However, as $\bar{y}_1$ depends on $\Delta$, the proof should be done considering four possible cases, i.e., $\bar{y}_1 \in \{1 \pm j, -1 \pm j\}$. Moreover, for each case of $\bar{y}_1 \in \{1 \pm j, -1 \pm j\}$, some complicated calculation is required to find $a$ satisfying $p_1<p_2$ through the triangle inequalities related to both $\theta$ and $\Delta$.

\begin{figure}[t]
    \centering
    \includegraphics[width=0.7\columnwidth]{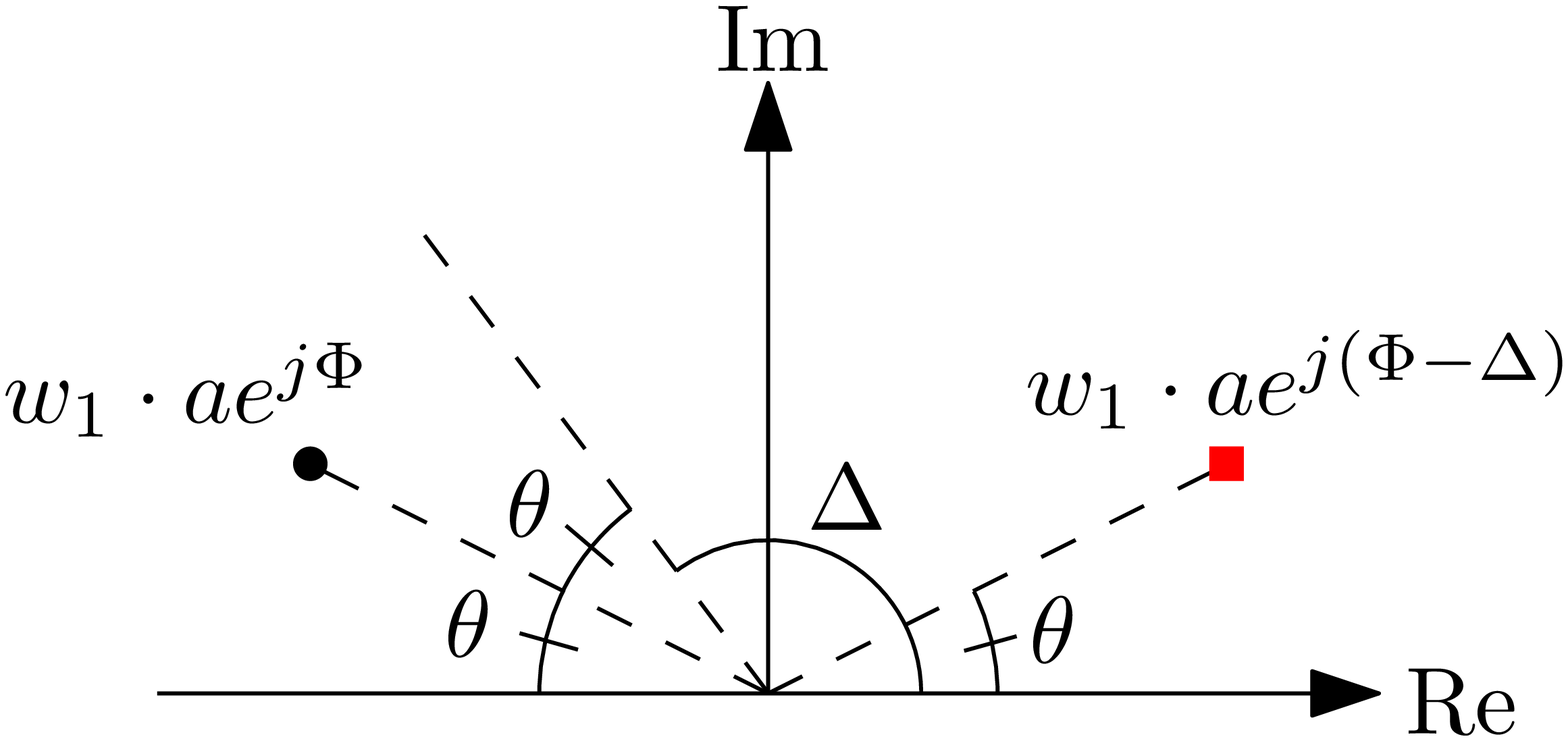}
    \caption{The relation between $\theta$ and $\Delta$ for $m=2$ in \eqref{eq:alignment}. By the symmetry, \eqref{eq:symmetry} holds.}
    \label{fig:Fig_Symmetry}
\end{figure}

To simplify the analysis, we choose $\theta$ (or $\Phi$, equivalently) to satisfy
\begin{equation}
    \theta + \Delta = m\pi/2 - \theta \text{ if } (m-1)\pi/2 \leq \Delta < m\pi/2, \label{eq:alignment}
\end{equation}
for $m \in \mathbb{Z}$.
Then,  $p_2$ is given as \eqref{eqn:p2} because $0<\theta \leq \pi/4$, and $p_1$ can be shown to be given as follows for all $\Delta$: 
\begin{multline}
    p_1 = P_{Y_1|X}\left(1+j \middle| ae^{j(\Phi-\Delta)}\right) \\ = P_{Y_1|X}\left( 1+j \middle| ae^{j(\theta-\angle w_1 ) } \right). \label{eq:symmetry}
\end{multline}
The above equation implies that we can treat $\bar{y}_1$ as $1+j$ by regarding the input as $ae^{j(\theta-\angle w_1 )}$, and do not need to prove the theorem separately for all ${\bar{y}_1 \in \{ 1 \pm j, - 1 \pm j\}}$. 
This can be understood graphically, as in \mbox{Fig. \ref{fig:Fig_Symmetry}}. In this figure, the real and imaginary axes correspond to the threshold lines of the one-bit ADCs at the legitimate receiver, and $\theta + \Delta \in (\pi/2,\pi)$. Therefore, $\bar{y}_1 = -1+j$.
The crossover probability $p_1$ is the probability that $Y_1=-1+j$ when the black circle plus $N_1$ is the input of the quantization.
Because $N_1$ is the circular symmetric Gaussian noise, $p_1$ is equal to the probability that $Y_1=1+j$ when the red square plus $N_1$ is the input of the quantization. Similar interpretation is available for all $m \in \mathbb{Z}$.

Now, because $\theta$ is the only phase to consider to analyze $p_1$ and $p_2$ in \eqref{eqn:p2} and \eqref{eq:symmetry}, respectively, the condition for $a$ to satisfy $p_1<p_2$ can be found easily.

A precise proof for Theorem \ref{thm_achievability} is given in the following. 
\begin{IEEEproof}[Proof of Theorem \ref{thm_achievability}]
Let us first prove the theorem for the real channel with one-bit ADC at the legitimate receiver and $k$-point ADC at the eavesdropper, and then generalize it for the complex channel. 
\\ \indent 1) Real channel
\\ \indent The input-output relationships for a real channel are given as 
\begin{equation}
    Y_1 = \text{sgn}(w_1 X + N_1),
\end{equation}
\begin{equation}
    Y_2 = \mathcal{Q}_k (w_2 X + N_2),
\end{equation}
where $N_i \sim \mathcal{N}(0,1)$.
Let $P_X$ be a binary distribution given as  
\begin{equation}
    P_X(\text{sgn}(w_2)\cdot a) = \phi, \; P_X(\text{sgn}(w_2)\cdot b) = 1-\phi, \label{eq:binary_dist}
\end{equation}
for some $a,b \in \mathbb{R}$, and $\phi \in (0,1)$.

Let us first derive the limit values of the mutual information terms when $b$ tends to infinity:  
\begin{align}
    &\lim\limits_{b\rightarrow\infty} I(X;Y_2) =  H(X) - \lim\limits_{b\rightarrow\infty}H(X|Y_2)
    \\& = h(\phi) - \lim\limits_{b\rightarrow \infty} P_{Y_2}(y_k) \cdot H(X|Y_2=y_k) \label{eq:thm1_1}
    \\& = h(\phi) - (1-\phi + \phi p_2) \cdot h\left(\frac{1-\phi}{1-\phi + \phi p_2 }\right)
    \\& = h(\phi(1-p_2)) - \phi h(p_2), \label{eq:thm1_2}
\end{align}
where
\begin{equation}
    p_2 = P_{Y_2|X}(y_k|\text{sgn}(w_2)\cdot a) = Q(q_{k-1}- |w_2| a). \label{eq:p2}
\end{equation}
Here, \eqref{eq:thm1_1} follows from the continuity of $P_{Y_2|X}(\cdot|\mathrm{sgn}(w_2)b)$ in $b$, $\lim\limits_{b\rightarrow \infty} P_{X|Y_2}(\text{sgn}(w_2)b|y_l)=0$ when $l<k$, and \eqref{eq:thm1_2} is from the following equations: 
\begin{multline}
    H(\phi p, 1-\phi, \phi-\phi p) = h(\phi) + \phi h(p) \\ = h(\phi(1-p)) + (1-\phi +\phi p) h\left( \frac{1-\phi}{1-\phi + \phi p} \right).
\end{multline}
Similarly, we can derive 
\begin{equation}
    \lim\limits_{b\rightarrow\infty} I(X;Y_1) = h \left( \phi(1-p_1) \right) - \phi h \left( p_1 \right),
\end{equation}
where
\begin{equation}
    p_1 = P_{Y_1|X}(\text{sgn}(w_1 w_2)|\text{sgn}(w_2)\cdot a) = Q(-|w_1| a). \label{eq:p1}
\end{equation}
Then, we get
\begin{equation}
   \lim\limits_{b \rightarrow \infty} R_s(P_X) = f_{\phi}(p_1) - f_{\phi}(p_2), \label{eq:Zchannel_difference}
\end{equation}
where ${f_\phi(p) := h (\phi(1-p)) - \phi h (p)}$.

The function $f_\phi (p)$ corresponds to the mutual information between input and output of the Z-channel with crossover probability $p$ when the input follows $\mbox{Bern}(\phi)$. Thus, it decreases in $p$.
\begin{lemma}\label{lem:f_decreasing}
For $0 < \phi < 1$, $f_\phi(p)$ is decreasing on ${p\in(0,1)}$.
\end{lemma}

The above lemma can be proved easily by checking the derivative of $f_\phi (p)$.
Then, \eqref{eq:p2}, \eqref{eq:p1}, \eqref{eq:Zchannel_difference}, Lemma \ref{lem:f_decreasing}, and the monotonicity of $Q(x)$ implies that $\lim\limits_{b \rightarrow \infty} R_s(P_X)>0$ if
\begin{equation}
    (|w_2|-|w_1|)a>q_{k-1}. \label{eq:real_proof}
\end{equation}
Because $|w_1| \neq |w_2|$, there exists $a$ satisfying \eqref{eq:real_proof}.
Now, due to the continuity of $R_s(P_X)$ in $b$, we conclude that there exist $\phi,a$, and $b<\infty$ such that $R_s(P_X)>0$. 

2) Complex channel
\\ \indent Let $P_X$ be a binary distribution 
\begin{equation}
    P_X(ae^{j\Phi}) = \phi, \; P_X(be^{j\Phi}) = 1-\phi, \label{eq:C_binary_dist}
\end{equation}
$0<\phi<1$, and $a,b,\Phi \in \mathbb{R}$. For a simple parameterization, let ${\theta := \Phi + \angle w_2}$, and $\Delta := \angle w_1 - \angle w_2$.
Then, the transition probabilities \eqref{eq:tran_p_11}-\eqref{eq:tran_p_2} can be represented as
\begin{multline}
    P_{Y_1|X}( 1 \pm j|xe^{j\Phi}) = Q( - |w_1|x\cos(\theta+\Delta)) \\ \cdot Q( \mp |w_1|x\sin(\theta+\Delta)),
\end{multline}
\begin{multline}
    P_{Y_1|X}(- 1 \pm j|xe^{j\Phi}) = Q( |w_1|x\cos(\theta+\Delta)) \\ \cdot Q( \mp |w_1|x\sin(\theta+\Delta)),
\end{multline}
\begin{multline}
    P_{Y_2|X}(y_{\mathrm{R},i}+j\cdot y_{\mathrm{I},l}|xe^{j\Phi})
    \\= \left(Q\left( q_{\mathrm{R},i-1}-|w_2|x\cos\theta \right) - Q\left( q_{\mathrm{R},i}-|w_2|x\cos\theta \right) \right)
    \\ \cdot \left(Q\left( q_{\mathrm{I},l-1}-|w_2|x\sin\theta \right) - Q\left( q_{\mathrm{I},l}-|w_2|x\sin\theta \right) \right),
\end{multline}
where $x \in \{a,b\}$.

For given $\Delta$, choose $\theta$ (or $\Phi$, equivalently) such that \eqref{eq:alignment} is satisfied.  Similar to the real case, let us evaluate the limit values of the mutual information terms as $b$ tends to infinity. 

Because $0<\theta\leq \pi/4$ under the condition \eqref{eq:alignment}, we have 
\begin{equation}
    \lim\limits_{b\rightarrow\infty}P_{X|Y_2}(be^{j\Phi}|y_{\mathrm{R},i}+j\cdot y_{\mathrm{I},l}) = 0,
\end{equation}
for $i\neq k_{\mathrm{R}}$ or $l \neq k_{\mathrm{I}}$. Thus, the limit value of $I(X;Y_2)$ is given as 
\begin{equation}
    \lim\limits_{b\rightarrow \infty} I(X;Y_2) = f_\phi(p_2),
\end{equation}
where
\begin{multline}
    p_2 = Q\left( q_{\mathrm{R},k_{\mathrm{R}}-1}-|w_2|a\cos\theta \right)
    \\ \cdot Q\left( q_{\mathrm{I},k_{\mathrm{I}}-1}-|w_2|a\sin\theta \right). \label{eq:x2}
\end{multline}
For the limit value of $I(X;Y_1)$, let $\bar{y}_1$ be the point such that
\begin{equation}
    \lim\limits_{b\rightarrow \infty} P_{Y_1|X}(\bar{y}_1|be^{j\Phi})=1.
\end{equation}
Because $m\pi/2-\pi/4\leq\theta+\Delta<m\pi/2$,
\begin{equation}
    \bar{y}_1 = \sqrt{2}\left(\cos \left( \frac{m \pi}{2} - \frac{\pi}{4} \right) + j \cdot \sin \left( \frac{m \pi}{2} - \frac{\pi}{4} \right) \right).
\end{equation}
Then, we have 
\begin{equation}
    \lim\limits_{b\rightarrow \infty} I(X;Y_1) = f_\phi(p_1),
\end{equation}
where $p_1$ is given as 
\begin{align}
    p_1 &= P_{Y_1|X}(\bar{y}_1|ae^{j\Phi})
    \\& = Q\left( -\sqrt{2}|w_1|a\cos \left( \frac{m \pi}{2} - \frac{\pi}{4} \right)\cos \left( \frac{m \pi}{2} - \theta \right)  \right)
    \\& \quad\quad \cdot Q\left( -\sqrt{2}|w_1|a\sin \left( \frac{m \pi}{2} - \frac{\pi}{4} \right)\sin \left( \frac{m \pi}{2} - \theta \right)  \right) \nonumber
    \\& = Q\left(-|w_1|a\left(t  \cos \theta  + (1-t)  \sin \theta\right)  \right)
    \\& \quad\quad \cdot Q\left(-|w_1|a\left((1-t)  \cos \theta  + t  \sin \theta\right)  \right) \nonumber
    \\& = Q(-|w_1|a \cos \theta) Q(-|w_1|a \sin \theta), \label{eq:x1}
\end{align}
where $t = \cos (m \pi /2)^2$, and the last equality follows from $t \in \{0,1\}$.

By Lemma \ref{lem:f_decreasing}, $\lim\limits_{b\rightarrow \infty} R_s(P_X) = f_\phi (p_1) - f_\phi (p_2)>0$ if ${p_1<p_2}$.
From \eqref{eq:x2}, \eqref{eq:x1}, and the monotonicity of $Q(\cdot)$, ${p_1<p_2}$ holds if 
\begin{equation}
    \begin{cases} q_{\mathrm{R},k_{\mathrm{R}}-1}-|w_2|a\cos\theta &< -|w_1|a \cos \theta \\ q_{\mathrm{I},k_{\mathrm{I}}-1}-|w_2|a\sin\theta &< -|w_1|a \sin \theta \end{cases},
\end{equation}
or equivalently,
\begin{equation}
    \begin{cases} q_{\mathrm{R},k_{\mathrm{R}}-1} &< (|w_2|-|w_1|)a \cos \theta \\ q_{\mathrm{I},k_{\mathrm{I}}-1} &< (|w_2|-|w_1|)a \sin \theta \end{cases}. \label{eq:complex_achieve_a}
\end{equation}
Because $\cos \theta, \; \sin\theta >0$, and $|w_2|-|w_1| \neq 0$, there exists $a$ which satisfies the above inequalities. Now, by the continuity of $R_s(P_X)$ in $b$, we conclude that  there exist $\phi,a$ and $b<\infty$ such that $R_s(P_X)>0$.
\end{IEEEproof}

Theorem \ref{thm_achievability} is generalized for a channel with arbitrary finite-resolution ADCs at the legitimate receiver as follows. 
\begin{corollary}\label{cor:general}
For a Gaussian wiretap channel with finite-resolution ADCs at both the legitimate receiver and the eavesdropper, there exists $P_X$ such that $R_s(P_X)>0$ and $\mathbb{E}[|X|^2] < \infty$ whenever $|w_1|\neq |w_2|$.
\end{corollary}

\begin{IEEEproof}
Let us first assume arbitrary 2-point ADCs at the legitimate receiver with threshold points $c_{\mathrm{R}}$ and $c_{\mathrm{I}}$ for the real and imaginary parts, respectively, and finite-resolution ADCs at the eavesdropper with threshold points ${\mathbf{q}_{\mathrm{R}} = (q_{\mathrm{R},1}, \cdots , q_{\mathrm{R},k_\mathrm{R}-1})}$ and ${\mathbf{q}_{\mathrm{I}} = (q_{\mathrm{I},1}, \cdots , q_{\mathrm{R},k_\mathrm{I}-1})}$ for the real and imaginary parts, respectively. 
Now, let ${\tilde{X}= X + c_{\mathrm{R}}/w_1 + j \cdot c_{\mathrm{I}}/w_1}$. Note that $\tilde{X}$ is a translated version of $X$. Hence it can be shown that if $P_X$ achieves a positive secrecy rate $R_s>0$ when the symmetric one-bit ADCs are at the legitimate receiver and finite-resolution ADCs with the threshold points $\mathbf{q}_{\mathrm{R}} - \mathcal{R}\left(\frac{w_2}{w_1}(c_{\mathrm{R}} + j \cdot c_{\mathrm{I}}) \right)$ and $\mathbf{q}_{\mathrm{I}} - \mathcal{I}\left(\frac{w_2}{w_1}(c_{\mathrm{R}} + j \cdot c_{\mathrm{I}}) \right)$ are at the eavesdropper, then $P_{\tilde{X}}$ achieves the same secrecy rate $R_s$ for the initially assumed channel. Therefore, we can conclude that a positive secrecy rate is achievable when arbitrary 2-point ADCs are at the legitimate receiver and finite-resolution ADCs are at the eavesdropper. 

Now, suppose finite-resolution ADCs at the legitimate receiver with threshold points  $\mathbf{c}_{\mathrm{R}}$ and $\mathbf{c}_{\mathrm{I}}$, which contain $c_{\mathrm{R}}$ and $c_{\mathrm{I}}$, respectively. Then, by the data processing inequality, a positive secrecy rate is also achievable. Because arbitrary $\mathbf{c}_{\mathrm{R}}$, $\mathbf{c}_{\mathrm{I}}$, $\mathbf{q}_{\mathrm{R}}$, and $\mathbf{q}_{\mathrm{I}}$ are assumed, we conclude that a positive secrecy rate is achievable as long as $|w_1|\neq |w_2|$ for arbitrary finite-resolution ADCs at both the legitimate receiver and at the eavesdropper.
\end{IEEEproof}

\begin{remark}
For $|w_2| \approx 0$, intuitively the achievability scheme used in the proof with almost all pairs of $(a,b)$ will achieve a positive secrecy rate, but \eqref{eq:complex_achieve_a} says that it is guaranteed only for $a$ less than some constant even if $|w_2|=0$. This seemingly counter-intuitive result is because \mbox{${\lim\limits_{b\rightarrow\infty}\lim\limits_{|w_2|\rightarrow 0} R_s(P_X) \neq \lim\limits_{|w_2|\rightarrow 0}\lim\limits_{b\rightarrow\infty} R_s(P_X)}$} in general. 
Taking the limit $|w_2| \rightarrow 0$ first means  ignoring the eavesdropper channel first. In this case, ${\lim\limits_{|w_2|\rightarrow 0} R_s(P_X)=I(X;Y_1)}$, so every pair of $(a,b)$ achieves a positive secrecy rate.
On the other hand, taking the limit $b\rightarrow \infty$ first means applying the Z-channel intuition first. In this case, as long as $|w_2|>0$, $\lim\limits_{b\rightarrow \infty} R_s(P_X)>0$ is not guaranteed for some $a$.
\end{remark}

\section{On the Optimal Distributions for the Wyner Code} \label{sec:optimal_support}
In Section \ref{sec:achievability}, it is shown that a binary distribution $P_X$ such that two input points have the same phase if $|w_1|<|w_2|$ and otherwise the opposite phase achieves a positive secrecy rate, when the one-bit ADCs are at the legitimate receiver. The following theorem states that  for a real Gaussian wiretap channel with one-bit ADCs at both the legitimate receiver and the eavesdropper, this choice of the phase is a necessary condition for the optimal distributions $P_X^*$ for the Wyner code, i.e.,
\begin{equation}
    R_s(P_X^*) = \sup\limits_{P_X:\mathbb{E}[X^2]\leq J} R_s(P_X). \label{eq:P_star}
\end{equation}

\begin{theorem}\label{Thm:optimal_support}
For a real Gaussian wiretap channel with one-bit ADCs, if $|w_1|<|w_2|$,
\begin{equation}
    \mathcal{S}(P_X^*) \subset \mathbb{R}_+ \text{ or } \mathcal{S}(P_X^*) \subset \mathbb{R}_-. \label{eq:Thm4.1_1}
\end{equation}
If $|w_1|>|w_2|$,
\begin{equation}
    \mathcal{S}(P_X^*) \not\subset \mathbb{R}_+ \setminus \{0\} \text{ and } \mathcal{S}(P_X^*) \not\subset \mathbb{R}_- \setminus \{0\}. \label{eq:Thm4.1_2}
\end{equation}
\end{theorem}

The following lemma plays a key role to prove the above theorem. 
\begin{lemma} \label{Lem:folded}  
For a real Gaussian wiretap channel with one-bit ADCs at both the legitimate receiver and the eavesdropper, if $|w_1|<|w_2|$,
\begin{equation}
 R_s \left( P_{|X|} \right) \geq R_s(P_X), \label{eq:Lem5.2_1}
\end{equation}
 and if $|w_1|>|w_2|$,
\begin{equation}
 R_s \left( P_{|X|} \right) \leq R_s(P_X), \label{eq:Lem5.2_2}
\end{equation}
 for all $P_X$ which satisfies ${\mathbb{E}[X^2]<\infty}$. If $X$ is not a constant, then each equality holds if and only if $\mathcal{S}(P_X) \subset \mathbb{R}_+$ or ${\mathcal{S}(P_X) \subset \mathbb{R}_-}$.
\end{lemma}
\begin{IEEEproof}[Proof of Lemma \ref{Lem:folded}]
First note that the sign of $w_i$ does not affect the mutual information terms as follows: 
\begin{align}
    I(X;& Y_i) = h(\mathbb{E}[Q(w_i X)]) - \mathbb{E}[h(Q(w_i X))]
    \\&= h(1-\mathbb{E}[Q(-w_i X)]) - \mathbb{E}[h(1-Q(-w_i X))]
    \\&= h(\mathbb{E}[Q(-w_i X)]) - \mathbb{E}[h(Q(-w_i X))].
\end{align}
Therefore, we assume $w_1$, $w_2 > 0$ without loss of generality.
To prove the lemma, it is sufficient to consider the sign of ${R_s(P_{|X|})-R_s(P_X)}$. 
Both the distributions, $P_X$ and $P_{|X|}$, induce the same conditional entropies because 
\begin{align}
    &H(Y_i|X) = \mathbb{E}[h(Q(w_i X))]
    \\& =  \mathbb{E}[h(Q(w_i X));X>0] + \mathbb{E}[h(Q(w_i X));X \leq 0]
    \\& = \mathbb{E}[h(Q(w_i X));X>0] + \mathbb{E}[h(Q(-w_i X));X \leq 0]
    \\& = \mathbb{E}[h(Q(w_i |X|))]
     \\& = \mathbb{E}_{X\sim P_{|X|}}[h(Q(w_i X)].
\end{align}
Therefore, the difference becomes
\begin{align}
    R_s(P_{|X|}) &- R_s(P_X) \nonumber
    \\& = h(\mathbb{E} [Q(w_1|X|) ]) - h(\mathbb{E}[Q(w_1 X)])
    \\& \quad - \left\{ h(\mathbb{E} [Q(w_2 |X|) ]) - h(\mathbb{E}[Q(w_2 X)])  \right\} \nonumber
    \\& = F(w_1) - F(w_2),
\end{align}
where $F(w) := h(\mathbb{E} [Q(w|X| )]) - h(\mathbb{E}[Q(w X)])$.

Let us define $c(w)$ and $d(w)$ as
\begin{align}
    c(w) &= \frac{1}{2} \left( \mathbb{E} [Q(w|X|)] + \mathbb{E}[Q(wX)] \right)
    \\&= \mathbb{E} [Q(wX);X>0] + \frac{1}{2}  \mathbb{E} [Q(-wX);X \leq 0]
    \\& \quad + \frac{1}{2}\mathbb{E}[Q(wX);X\leq 0] \nonumber
    \\& = \mathbb{E} [Q(wX);X>0] + \frac{1}{2} P_X(\mathbb{R_-}),
\end{align}
and
\begin{align}
    d(w) &= \frac{1}{2} \left( \mathbb{E}[Q(wX)] - \mathbb{E} [Q(w|X| )] \right)
    \\&= \frac{1}{2} \left( \mathbb{E} [Q(wX);X\leq0] - \mathbb{E}[Q(-wX);X\leq 0] \right)
    \\& = \mathbb{E} [Q(wX);X\leq0] - \frac{1}{2} P_X(\mathbb{R_-}).
\end{align}
Then, the function $F(\cdot)$ can be represented as
\begin{equation}
    F(w) = h(c(w)-d(w)) - h(c(w)+d(w)).
\end{equation}

By the monotonicity of $Q(\cdot)$, it can be checked that $c(w)$ is non-increasing, $d(w)$ is non-decreasing, and ${0 \leq d(w) <c(w) \leq 1/2}$ for $w>0$. The strictly inequality $d(w)<c(w)$ follows from $\mathbb{E}[X^2]<\infty$. Furthermore, the following lemma can be proved easily by deriving the partial derivatives.
\begin{lemma}
For $0<d<c<1/2$, $h(c-d)-h(c+d)$ is strictly increasing in $c$ for fixed $d$, and strictly decreasing in $d$ for fixed $c$.
\end{lemma}

Hence, from the above lemma and the  monotonicity of $c(w)$ and $d(w)$, it follows that $F(w)$ is decreasing in $w$, which proves  $\eqref{eq:Lem5.2_1}$ and $\eqref{eq:Lem5.2_2}$. 

Now let us check the equality conditions in $\eqref{eq:Lem5.2_1}$ and $\eqref{eq:Lem5.2_2}$. Suppose $X$ is not a constant. Then, at least one of $c(w)$ and $d(w)$ is strictly monotonic in $w>0$. Hence, $F(w)$ remains constant if and only if $c(w) = 1/2$ or $d(w) = 0$ for all $w>0$. The equality conditions follow from the fact that $c(w)=1/2$ if and only if $\mathcal{S}(P_X) \subset \mathbb{R_-}$, and $d(w)=0$ if and only if $\mathcal{S}(P_X) \subset \mathbb{R}_+$.
\end{IEEEproof}
Now we are ready to prove Theorem \ref{Thm:optimal_support}.
\begin{IEEEproof}[Proof of Theorem \ref{Thm:optimal_support}]
Because $\sup\limits_{P_X:\mathbb{E}[X^2]\leq J}R_s(P_X) > 0$ by Theorem \ref{thm_achievability}, $X$ is not a constant. Since $P_X^*$ is the optimal distribution, the inequality \eqref{eq:Lem5.2_1} should be equality when $|w_1|<|w_2|$ for $P_X^*$. Therefore, \eqref{eq:Thm4.1_1} follows from the equality condition in Lemma \ref{Lem:folded}.

For $|w_1|>|w_2|$, suppose $\mathcal{S}(P_X^*) \subset \mathbb{R}_+ \setminus \{0\}$. Choose $x \in \mathcal{S}(P_X^*)$ which is not the maximum, and define $P_X'$ as
\begin{equation}
    P_X'(A) = P_X^*( -(A \cap [-x,0)) ) + P_X^*(A \cap (x,\infty)),
\end{equation}
for every Borel set $A \subset \mathbb{R}$. Then, $P_{|X|}' = P_X^*$, and $\mathcal{S}(P_X')$ is not the subset of $\mathbb{R}_+$ or $\mathbb{R}_-$. By the equality condition in Lemma \ref{Lem:folded}, $R_s(P_X^*) < R_s(P_X')$, which contradicts to the optimality of $P_X^*$. Therefore, $\mathcal{S}(P_X^*) \not\subset \mathbb{R}_+ \setminus \{0\}$. Similarly, it can be shown that $\mathcal{S}(P_X^*) \not\subset \mathbb{R}_- \setminus \{0\}$.
\end{IEEEproof}

\section{Conclusion and Discussions}\label{sec:discussion}
In this paper, we showed that a positive secrecy rate is always achievable for a Gaussian wiretap channel with ADCs, as long as the channel gains are not the same. For the achievability, we first showed the achievability for the case when the one-bit ADCs are at the legitimate receiver. For such a case, we employed binary channel inputs with the same phase or opposite phase depending on the channel conditions, and analyzed the achievable rate by approximating the resultant channels to Z-channels. Then the result was generalized  for the channel with arbitrary finite-resolution ADCs at the legitimate receiver through some simple manipulation. Moreover, for a real Gaussian wiretap channel with one-bit ADCs, such phase condition used for the achievability was shown to satisfy a necessary condition for the optimal input distribution for the Wyner code.

The Wyner code we considered in this paper can be applied to practical digital communication systems.  As we mentioned in Section \ref{subsec:prev}, the polar code in \cite{MV11} for wiretap channel achieves $R_s(P_X)$ for binary uniform input distribution $P_X$. As our achievability holds for uniform distribution, i.e., ${\phi=1/2}$ in \eqref{eq:binary_dist}, the code in \cite{MV11} with the modulation which we used in the proof of Theorem \ref{thm_achievability} can be used for a secure communication system in practice.

A potential future work of interest would be to characterize the exact secrecy capacity and the optimal distributions which achieve it.
In Section \ref{subsec:prev}, we mentioned that the solution of \eqref{eq:Degraded_Cs} would be helpful for analyzing the secrecy capacity. 
Our expectation comes from the fact that
\begin{multline}
    I(U;Y_1)-I(U;Y_2) = I(X;Y_1)-I(X;Y_2)
    \\ + \left[ I(X;Y_2|U) - I(X;Y_1|U) \right],
\end{multline}
for $U-X-(Y_1,Y_2)$, as in \cite{cyclic_symmetry}. Now, let $P_X^*$ denote a distribution achieving \eqref{eq:Degraded_Cs} and $\{\tilde{P}_X^i: i=1,2,\cdots \}$ denote the set of distributions achieving $\sup\limits_{P_X}I(X;Y_2)-I(X;Y_1)$. If $P_X^*$ can be represented as a weighted sum of $\tilde{P}_X^i$'s, i.e., there exists $\{\theta_i\}$ such that $\theta_i \geq 0, \sum\limits_i \theta_i = 1$ and 
\begin{equation}
    P_X^* = \sum\limits_i \theta_i \tilde{P}_X^i,
\end{equation}
the secrecy capacity is given by
\begin{multline}
    C_s = \max\limits_{P_X} [I(X;Y_1) - I(X;Y_2)] \\ + \max\limits_{P_X} [I(X;Y_2)-I(X;Y_1)]. 
\end{multline}
Therefore, the characterization of the optimal distributions which achieve \eqref{eq:Degraded_Cs} can be used to find the secrecy capacity.

In this work, we provided a necessary condition for the optimal distributions which achieve \eqref{eq:Degraded_Cs}. One possible way to find more necessary conditions is to apply Karush-Kuhn-Tucker (KKT) conditions.
There have been some studies on necessary conditions for the optimal support using KKT conditions \cite{p2pADC,SMITH1971203,dytsodiscrete} for point-to-point channels, where the cardinality of the optimal support is bounded.
 The KKT conditions that used for point-to-point channels \cite[Theorem 10]{dytsodiscrete} can be modified for wiretap channels. Precisely, if $P_X^*$ is that of \eqref{eq:P_star}, there exists $\lambda \geq 0$ such that following conditions are satisfied:
\begin{equation}
    \lambda(\mathbb{E}_{X\sim P_X^*}[X^2]-J)=0,
\end{equation}
\begin{multline}
    i(x;P_X^*,P_{Y_1|X})-i(x;P_X^*,P_{Y_2|X})-\lambda (x^2-\mathbb{E}_{X\sim P_X^*}[X^2])
    \\ \leq R_s(P_X^*),
\end{multline}
for all $x \in \mathbb{R}$, and equality holds if $x \in \mathcal{S}(P_X^*)$.\footnote{In our channel, only necessity is guaranteed because $R_s(P_X)$ is not concave in $P_X$.}
We expect that it would be possible to find a stronger condition by combining KKT conditions and Theorem \ref{Thm:optimal_support} in our work.

\begin{appendices}
\section{}
For a complex Gaussian wiretap channel with one bit ADCs at both the legitimate receiver and the eavesdropper, let us show that the QPSK aligned to the legitimate channel achieves a positive secrecy rate whenever $|w_1|>|w_2|$. 

In this channel, the transition probabilities are given by
\begin{align}
     P_{Y_i|X}( 1 \pm j|x) &= Q(- \mathcal{R}(w_i x))Q(\mp \mathcal{I}(w_i x)),
    \\ P_{Y_i|X}(- 1 \pm j|x) &= Q( \mathcal{R}(w_i x))Q(\mp \mathcal{I}(w_i x)),
\end{align}
for $i=1,2$.
Let $P_X(x) = 1/4$ for $x\in \mathrm{QPSK}_J \cdot e^{-j \angle w_1}$, where
\begin{equation}
    \mathrm{QPSK}_J := \sqrt{\frac{J}{2}} \cdot \{1\pm j, -1 \pm j\}.
\end{equation}
Then, it can be shown that $H(Y_i)=2$ for $i=1,2$ from
\begin{equation}
    P_{Y_i|X}(y_i|x) = P_{Y_i|X}(y_i\cdot e^{j\cdot \pi/2}|x\cdot e^{j\cdot \pi/2}),
\end{equation}
for all $i\in\{1,2\}$, $x\in \mathbb{C}$, and $y_i \in \{1 \pm j, - 1 \pm j\}$.
Then, the mutual informations are given as
\begin{equation}
    I(X;Y_1) = 2\left( 1-h\left( Q\left( |w_1|\sqrt{\frac{J}{2}} \right) \right) \right), \label{eq:QPSK_I1}
\end{equation}
\begin{align}
    I(X;Y_2) &= 2-H(Y_2|X)
    \\& = 2-h\left( Q\left( |w_2| \sqrt{J} \cos(\Delta) \right)\right)
    \\& \quad\quad\quad - h\left( Q\left( |w_2|\sqrt{J}  \sin(\Delta) \right)\right), \nonumber
\end{align}
where $\Delta = \angle w_1 - \angle w_2$.
Because $h\left( Q\left(\sqrt{\cdot} \right) \right)$ is convex \cite{hqsqrt}, we have
\begin{equation}
    I(X;Y_2) \leq  2\left(1- h\left( Q\left( |w_2|\sqrt{\frac{J}{2}} \right) \right) \right). \label{eq:QPSK_I2}
\end{equation}
Therefore, from \eqref{eq:QPSK_I1} and \eqref{eq:QPSK_I2},
\begin{multline}
    R_s(P_X) \\ \geq  2\left( h\left( Q\left( |w_2|\sqrt{\frac{J}{2}} \right) \right) -h\left( Q\left( |w_1|\sqrt{\frac{J}{2}} \right) \right) \right) \\ >0,
\end{multline}
if $|w_1|>|w_2|$.
\end{appendices}

\bibliographystyle{ieeetr}
\bibliography{ref}

\end{document}